# Physics Informed Neural Network Enhanced Denoising for Atomic Resolution STEM Imaging

Z. Awan, J. Shabeer, U. Saleem, S. Mehmood, T. Qadeer

**Abstract**

Atomic resolution STEM images often suffer from noise due to low electron doses and instrument imperfections, hence it is challenging to obtain critical structural details required for material analysis. To address the problem, we propose a Physics-Informed Neural Network (PINN) framework for denoising STEM images. Our method integrates spectral fidelity, total variation, and brightness/contrast consistency losses to ensure the preservation of fine structures, smooth regions, and physical signal intensities, maintaining the structural integrity of the denoised images. Our proposed method effectively balances noise reduction with the preservation of atomic resolution details and complements existing methods, seeking to enhance the utility of STEM images in material characterization and analysis.

**Introduction**

Scanning Transmission Electron Microscopy (STEM) is an important tool in materials science for imaging the atomic-scale features with high spatial resolution. STEM plays a vital role in understanding material properties, from crystallographic structures to defect dynamics[i,ii,iii,iv,v]. However, the inherent trade-offs in imaging conditions, such as the need for low electron doses to minimize sample damage result in significant noise that may obscure important structural and compositional details[vi,vii,viii,ix,x]. Therefore, it is essential to develop effective denoising methods to maximize the utility of STEM images for analysis and interpretation.

Traditional denoising methods, such as Gaussian filters, Fourier filtering, or wavelet transforms, are often effective at reducing noise but may inadvertently blur critical details or introduce artifacts. Recent advances in deep learning have demonstrated significant promise in addressing such challenges[xi,xii,xiii]. Neural networks excel at capturing complex relationships in data, enabling them to remove noise while retaining features like sharp edges and subtle textural details. However, generic neural networks may struggle to ensure that the denoised outputs align with the physical properties of the system being studied.

In this work, we propose a Physics-Informed Neural Network (PINN) framework tailored for STEM image denoising. Our approach integrates domain-specific knowledge into the learning process by incorporating spectral fidelity, total variation, and brightness/contrast consistency losses. These physics-inspired constraints ensure that the denoised images retain critical features such as lattice patterns, defect structures, and smooth backgrounds, all of which are essential for accurate material characterization.

The spectral fidelity loss emphasizes the preservation of frequency-domain information, ensuring that fine structural details remain intact. Total variation loss encourages smoothness in uniform regions while preserving sharp transitions, such as those found at atomic boundaries or defects. Brightness and contrast consistency losses ensure that the denoised image reflects the physical intensity levels of the original STEM data, maintaining its scientific relevance.

**Methods:**

Synthetic STEM images were produced to mimic atomic resolution imaging for various atomic arrangements such as silicon, gallium nitride, aluminum gallium nitride, perovskites, graphene, and other systems of interest. The structural arrangements were derived from Crystallographic Information Files (CIFs) provided by the Burai 1.3 software, which offers standard examples of crystalline structures. Custom Python scripts processed the CIF data to generate diverse lattice structures, with atomic positions and intensities modeled to reflect characteristic features of the selected materials. Gaussian filtering was used to simulate the broadening effects of an electron beam in STEM imaging, while adjustments in intensity and contrast enhanced the realism of atomic scattering. Custom Python scripts used to process CIF files to simulate lattice structures.

To emulate real-world imaging conditions, synthetic STEM images were degraded using a Composite Degradation Noise (CDN) model, which integrates multiple noise types into a unified framework. Three primary noise types were considered: Gaussian noise, static noise, and speckle noise. The noise was systematically applied at different intensity levels to mimic a range of experimental conditions. This step bridges the gap between ideal simulations and practical challenges in experimental STEM imaging, making the data highly relevant for testing and validating denoising methods. The CDN is described in more detail below.

**Composite Degradation Noise (CDN)**

The CDN is mathematically expressed as follows:

$$I_{CDN}(x,y) = (I(x,y) + N_{gaussian}(x,y)) \cdot (1 + N_{speckle}(x,y))$$

where $I(x,y)$ denotes original image intensity, $N_{gaussian}(x,y)$ the Gaussian noise added to each pixel, and $N_{speckle}(x,y)$ introduces multiplicative variations characteristic of speckle noise. Static noise, however, is applied as a replacement operation on a fraction of randomly selected pixels during the Gaussian stage, disrupting localized intensities.

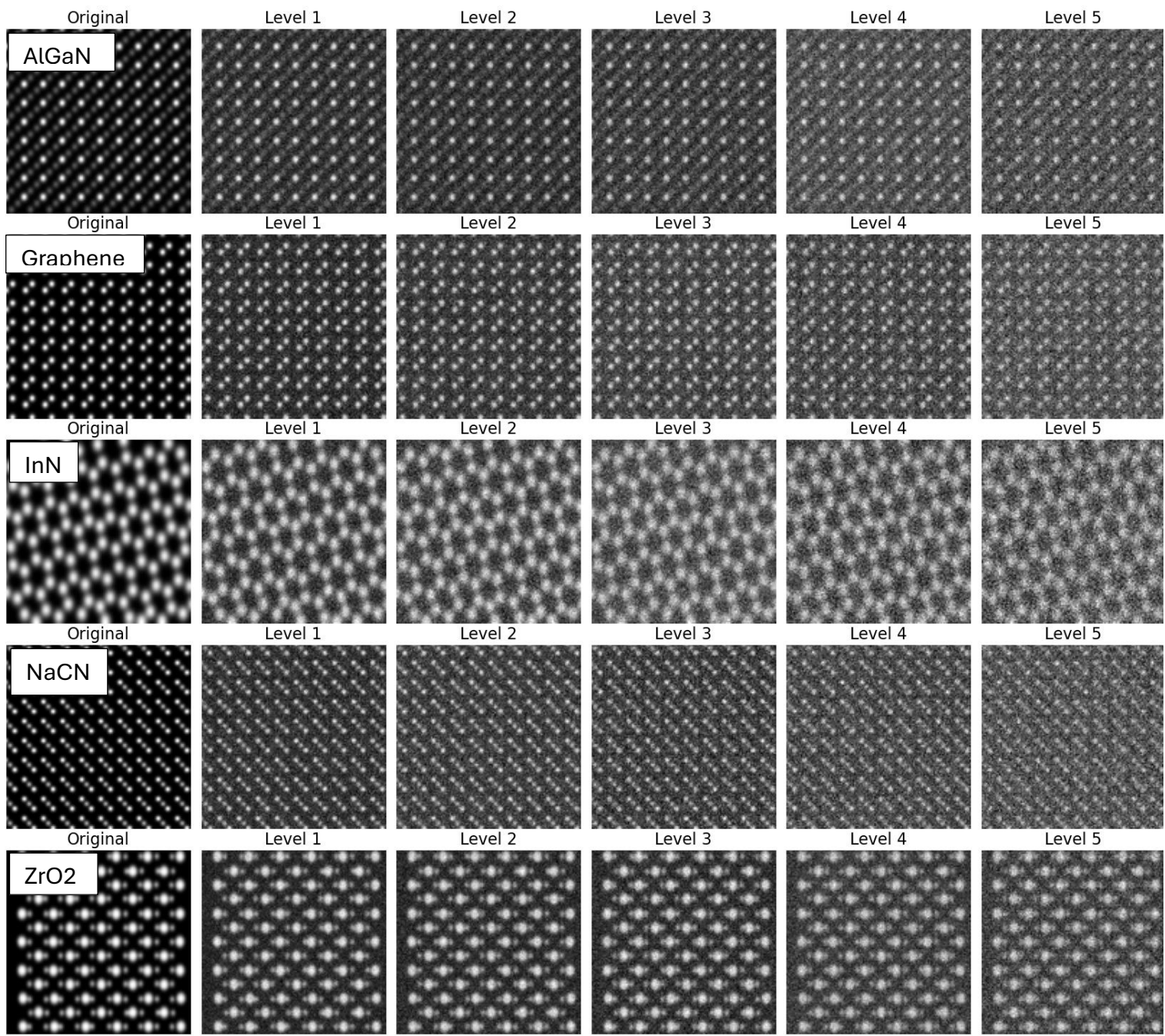

**Figure 1:** Clean and noised synthetic STEM images

Each of the noise components of CDN are mathematically described below

### Gaussian Noise

Gaussian noise accounts for electronic fluctuations and background disturbances and is modeled as

$$N_{gaussian}(x,y) \sim \mathbb{N}(0, \sigma^2_{gaussian}(x,y))$$

where $\sigma^2_{gaussian}$ controls the standard deviation. This noise is added to the original image $I(x,y)$ to form

$$I(x,y) = I(x,y) + N_{gaussian}(x,y)$$

### Static Noise:

Static noise introduces sharp intensity variations by replacing a subset of pixel values with random values sampled uniformly from the range [0,255]. This simulates sensor defects or data corruption

$$I_{static} = \begin{cases} Rand(0, 255), & if\ (x,y) \in S \\ I_{gaussian}(x,y), & otherwise \end{cases}$$

Here, $S$ denotes the set of randomly selected pixel coordinates, determined by the static intensity parameter $p_{static}$.

### Speckle Noise

Speckle noise introduces multiplicative distortions, simulating coherent interference patterns

$$N_{speckle}(x,y) \sim \mathbb{N}(0, \sigma^2_{speckle}(x,y))$$

It modulates the Gaussian and static-noise-affected image as

$$I(x,y) = I(x,y)\{1 + N_{speckle}(x,y)\}$$

In the CDN process, Gaussian noise is applied first to introduce global, small-scale fluctuations, followed by Static noise that replaces randomly selected pixel intensities and then the Speckle noise follows. Speckle noise introduces intensity dependent multiplicative distortions. The result of this composite noise is a realistic experimental degradation in STEM images.

Each of the CDN components is numerically expressed in table 1 defining each noise level. It is evident how each of the STEM images gets degraded with increasing noise level.

| Noise Level | Gaussian (Std Dev) | Static (Intensity) | Speckle (Variance) | Comments |
|---|---|---|---|---|
| 1 | 50 | 0.50 | 0.20 | Moderate Gaussian blur, 50% pixels randomly altered, mild speckle |
| 2 | 60 | 0.60 | 0.30 | Slightly stronger Gaussian noise, 60% static noise, moderate speckle |
| 3 | 70 | 0.70 | 0.40 | Increased Gaussian noise, 70% static noise, noticeable speckle |
| 4 | 80 | 0.80 | 0.50 | Heavy Gaussian blur, 80% random noise, strong speckle artifacts |
| 5 | 100 | 0.90 | 0.60 | Maximum Gaussian blur, 90% pixels altered, very pronounced speckle |

**Table 1:** Summary of various DCN components for each level of noise introduced to synthetic STEM images

**Physics Informed Neural Network**

Our Physics-Informed Neural Network Enhanced Denoising (PINNED) model is designed to reconstruct high-quality denoised images while preserving their structural and physical fidelity by embedding physics informed constraints into the training process. These constraints enable our model to follow known properties of the STEM imaging system and noise characteristics.

The training of PINNED model was performed on paired datasets of noisy and clean STEM images. As discussed before, the noisy images were generated by applying a composite degradation noise (CDN) model to high-quality images to simulate experimental defects commonly found in real STEM images. The model maps between noisy inputs and clean outputs to closely resemble the denoised reconstructions with clean images while removing the introduced noise.

The architecture of PINNED model is based on a deep convolutional encoder decoder network and graded feature extraction. The encoder gradually compresses the input image into a latent representation to learn on its features that are necessary for denoising. This latent representation is then processed through a decoder to reconstruct the image in reverse, using transposed convolutional layers to upscale the resolution. Each encoder and decoder block employs three successive convolutional layers with Leaky ReLU activations to ensure feature richness and non-linearity, enhancing the model's ability to capture subtle image features.

Mathematically, the PINNED model maps a noisy input image, $y$, to a denoised output, $\hat{x}$, through a parametric function $f_\theta$, defined by the neural network:

$$\hat{x} = f_\theta(y)$$

The noisy image $y$ is encoded to a latent representation $z$ by the encoder $Enc_{\theta E}$ via the operation defined by equation

$$z = Enc_{\theta E}(y)$$

followed by a decoding operation, in which the decoder $Enc_{\theta E}$ decodes the latent representation $z$ and reconstructs the image $\hat{x}$. The decoding operation is defined by equation

$$\hat{x} = Dec_{\theta D}(z)$$

The role of encoder is to generate a compact latent representation '$z$' that holds essential features of the input image while removing anything else as noise. This latent space is then expanded by the decoder to reconstruct the denoised image.

To make denoising process respect the physical properties of the images, PINNED uses a set of physics informed constraints within the loss function. These constraints enforce structural and spectral fidelity in image reconstruction. These constraints were made part of the Loss function and therefore are key components of the training process. Each of the components of loss function are described in more detail below.

**Total variation (TV) loss** penalizes any abrupt changes in the reconstructed image while preserving edges. Mathematically defined as

$$\mathcal{L}_{TV} = \sum_{i,j} \{|\hat{x}_{i+1,j} - \hat{x}_{i,j}| + |\hat{x}_{i,j+1} - \hat{x}_{i,j}|\}$$

**Spectral fidelity (SF) loss** is defined in the frequency domain to maintain consistency between the spectral characteristics of the noisy and clean images. It minimizes the difference in magnitudes of the Fourier transforms of the reconstructed and target images. It is mathematically defined by

$$\mathcal{L}_{SF} = \frac{1}{N} \sum_{k} ||\mathcal{F}(\hat{x})(k)| - |\mathcal{F}(x)(k)||$$

Where $\mathcal{F}$ represents the Fourier transform, $k$ denotes frequency components, and $N$ is the total number of pixels.

**Brightness and Contrast Consistency losses** were used to ensure that the reconstructed images retain global intensity and feature prominence like clean STEM images. These are defined as

$$\mathcal{L}_B = |\mu(\hat{x}) - \mu(x)|$$

where $\mu(\hat{x})$ and $\mu(x)$ are mean intensities of reconstructed and original image respectively. Similarly, the contrast consistency loss enforces standard deviation alignment between the predicted and target images, mathematically define as

$$\mathcal{L}_C = |\sigma(\hat{x}) - \sigma(x)|$$

where $\sigma(\hat{x})$ and $\sigma(x)$ are the intensity standard deviation of reconstructed and original images.

The **total loss function** was defined as a weighted sum of previously defined loss components and generic loss defined by mean squared difference (MSE) between original and reconstructed images as

$$\mathcal{L}_{total} = \mathcal{L}_{MSE} + \lambda_{TV}\mathcal{L}_{TV} + \lambda_{SF}\mathcal{L}_{SF} + \lambda_B\mathcal{L}_B + \lambda_C\mathcal{L}_C$$

This PINNED framework further uses hierarchical feature extraction in encoder decoder architecture that through convolutional layers. Each layer applies a series of convolutional operations.

$$z_l = \phi(W_l * z_{l-1} + b_l)$$

Where $W_l$ and $b_l$ are the weights and biases, '*' represents convolution, and $\phi$ is the Leaky ReLU activation function. The decoder mirrors this process through transposed convolutions for upscaling.

**Model Validation**

The validation of the Physics-Informed Neural Network Enhanced Denoising (PINNED) model was carried out rigorously using both synthetic and experimentally degraded STEM images. The validation process aimed to evaluate the model's ability to denoise images effectively while preserving critical structural and spectral features. This section outlines the methodology and results obtained during the validation phase.

**Validation Dataset**

The dataset used for validation included synthetic STEM images generated from Crystallographic Information Files (CIFs) representing various material systems, such as silicon, gallium nitride, aluminum gallium nitride, and perovskites. These images were subjected to the Composite Degradation Noise (CDN) model, creating paired datasets of noisy and clean images. The noisy images simulated real-world imaging artifacts, including Gaussian noise, static noise, and speckle noise, as described earlier. The paired dataset ensured that the validation process could quantify the model's performance in reconstructing the clean images.

In addition to synthetic data, experimentally degraded STEM images were included to assess the model's performance in realistic scenarios. These images exhibited diverse noise profiles resulting from varying imaging conditions, such as beam instabilities, detector sensitivity, and sample vibrations.

**Performance Metrics**

The model's performance was evaluated using a combination of quantitative and qualitative metrics to ensure a comprehensive validation:

1. **Peak Signal-to-Noise Ratio (PSNR)**
   PSNR measures the fidelity of the reconstructed image compared to the clean image. It is defined as:

   $$PSNR = 10 log_{10}(\frac{MAX^2}{MSE})$$

   where MAX is the maximum possible pixel value and MSE is the mean squared error between the clean and reconstructed images. A higher PSNR indicates better reconstruction quality.

2. **Structural Similarity Index (SSIM):**
   SSIM quantifies the structural similarity between the reconstructed and clean images, considering luminance, contrast, and structural information:

   $$SSIM = \frac{(2\mu_x\mu_y + C_1)(2\psi_{xy} + C_2)}{(\mu_x^2 + \mu_y^2 + C_1)(\psi_x^2 + \psi_y^2 + C_2)}$$

   where $\mu_x$ and $\mu_y$ are the mean intensities, $\psi_x^2$ and $\psi_y^2$ are variances, $\psi_{xy}$ is the covariance, and $C_1 \; and \; C_2$ are constants to stabilize the division. SSIM values close to 1 indicate high structural similarity.

3. **Fourier Spectrum Consistency (FSC):**
   FSC measures the consistency of the reconstructed image's frequency spectrum with that of the clean image. The metric evaluates the similarity between the magnitude spectra of the Fourier transforms of the two images

   $$FSC = \frac{\sum_k |\mathcal{F}(\hat{x})(k)| . |\mathcal{F}(x)(k)|}{\sqrt{\sum_k |\mathcal{F}(\hat{x})(k)|^2 . |\mathcal{F}(x)(k)|^2}}$$

where $\mathcal{F}$ represents the Fourier transform.

4. **Edge Preservation Index (EPI):**
   The EPI assesses the preservation of sharp edges in the reconstructed image by comparing gradients:

   $$EPI = \frac{\|\nabla \hat{x}\|_1}{\|\nabla x\|_1}$$

   Values close to 1 indicate excellent edge preservation.

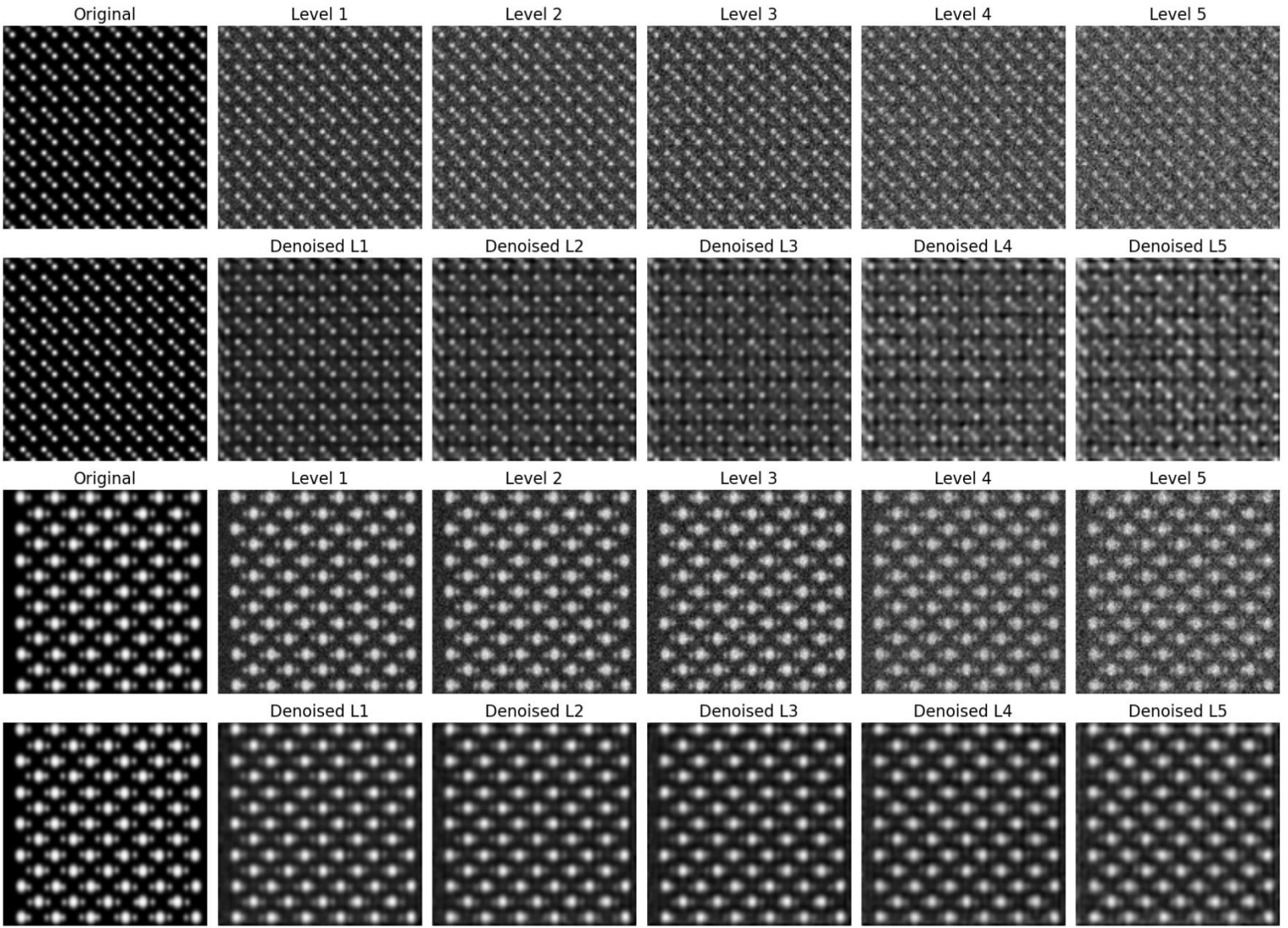

**Figure 2a:** Noisy and corresponding reconstructed images for training set

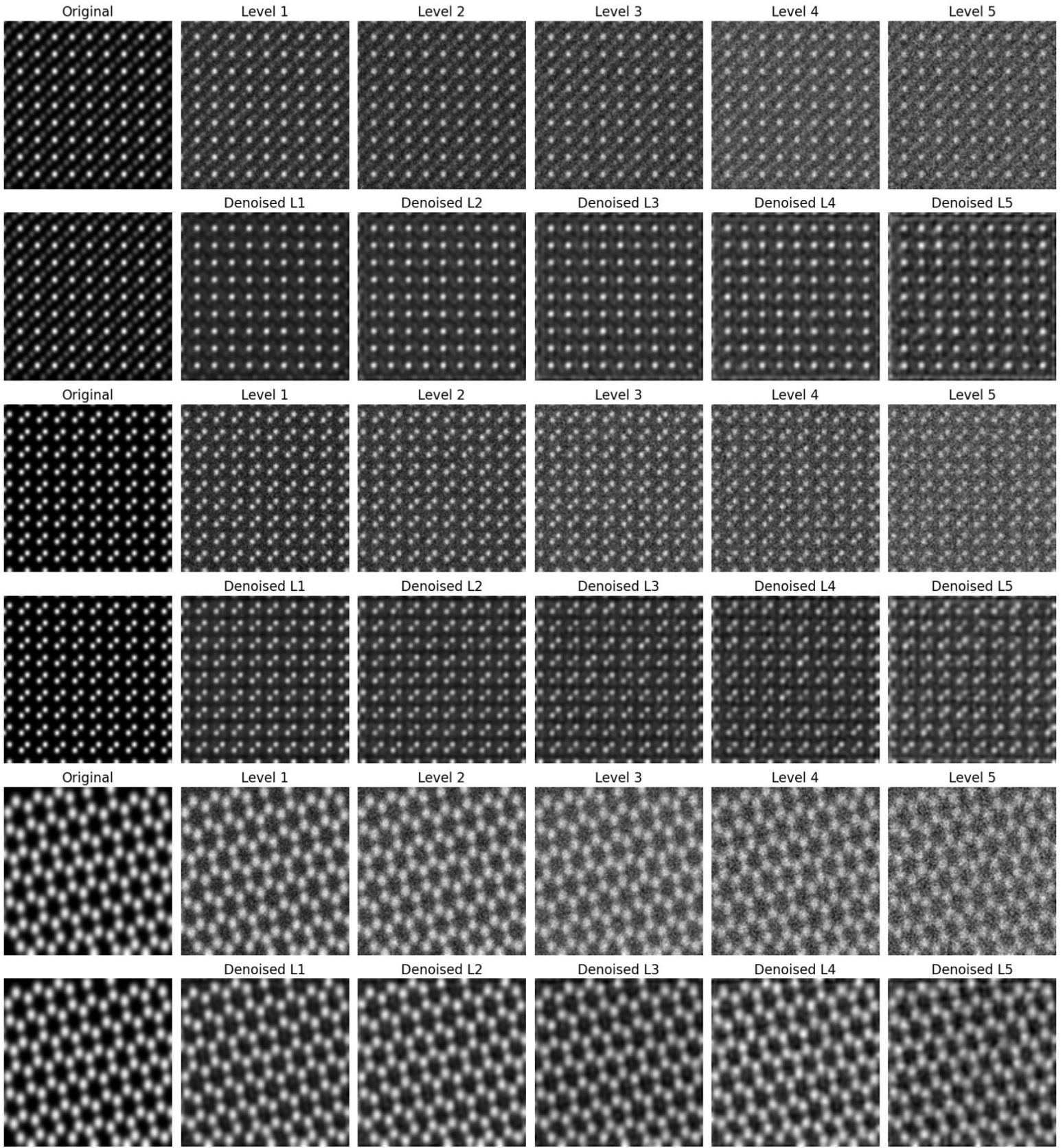

**Figure 2b:** Noisy and corresponding reconstructed images for training set

**Results and Analysis**

The PINNED model was trained on set of 5 STEM images for 500 epochs. The training loss went down to 0.0425. The model was was then validated on an external set of other STEM images that were never shown to the model during training phase.

**Training Set**

The original, noisy and reconstructed images from the training set are shown in Fig 2a and 2b. The performance evaluation of the Physics-Informed Neural Network Enhanced Denoising (PINNED) model is systematically conducted using four key metrics defined above. All four metrics are computed for images reconstructed from multiple noise levels, namely **Levels 1 through 5**, where each level is defined by a progressive increase in composite degradation noise comprising Gaussian, static, and speckle components. The results are visually displayed for all training images in Fig 3.

i. **Peak Signal to Noise Ratio (PSNR)**

The **PSNR** measures quantitative fidelity of the reconstructed images. As expected, PSNR decreases consistently with increasing noise levels across all samples. This decline indicates a proportional loss of reconstruction fidelity as noise intensity increases, reflecting the model's increasing challenge to accurately recover fine details. **InN** consistently demonstrates the highest PSNR across all noise levels, suggesting that its structural content and smooth spatial variations enable better reconstruction by the model. By contrast, **Graphene, ZrO2, AlGaN** and **NaCN** exhibit lower PSNR values, particularly at higher noise levels, indicating the model's reduced capacity to recover high-frequency details in these cases.

ii. **Structural Similarity Index (SSIM)**

The **SSIM** provides an assessment of the structural integrity between reconstructed and original images. Similar to PSNR, SSIM values for reconstructed images decline as noise levels increase. However, the rate of degradation in structural similarity is more pronounced, particularly for *NaCN* and *Graphene*. However, *InN* preserves higher SSIM values, maintaining structural consistency across noise levels.

iii. **Fourier Spectrum Consistency (FSC)**

The **FSC** evaluates the alignment of frequency components between reconstructed and original images. FSC values remain relatively stable for most images, particularly *Graphene* and *NaCN*, indicating that their frequency content is less disrupted by noise. However, *ZrO2* shows a sharp decline in FSC at higher noise levels, reflecting significant perturbation of spatial frequency information. This suggests that noise intensities at Level 4 and Level 5 introduce complex distortions in specific cases that challenge the model's ability to recover finer frequency details.

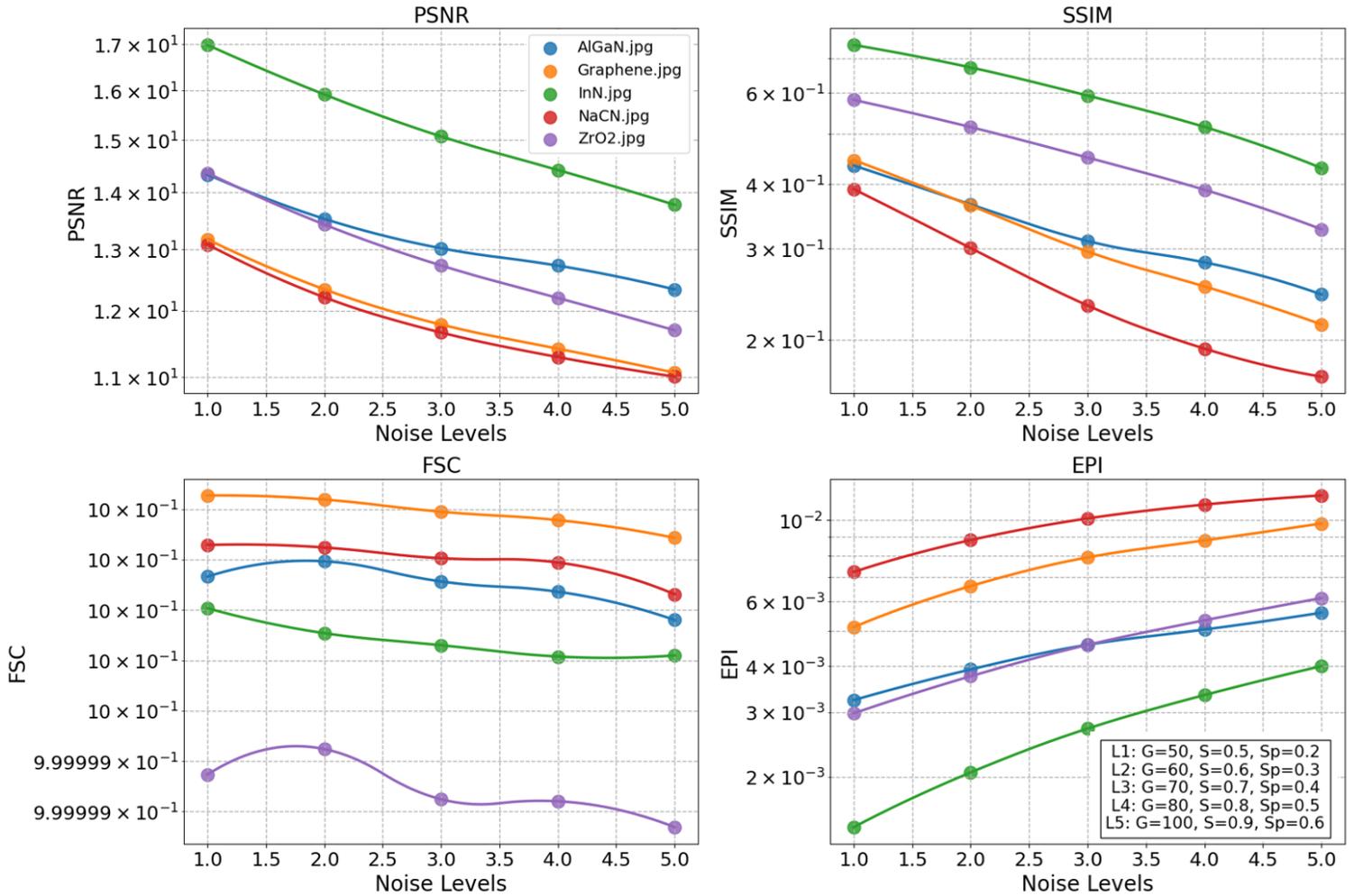

**Figure 3:** Performance evaluation metrics for all images in training set

### iv. Edge Preservation Index (EPI)

The **EPI** quantifies the preservation of edge features, which are critical for identifying structural details. EPI values exhibit a gradual increase with higher noise levels for all images, indicative of the added challenges in edge recovery as the noise becomes more severe. Among the samples, *NaCN* and *Graphene* maintain the highest EPI values, while *InN* shows relatively lower values, reflecting differences in the edge content and texture complexity of each sample.

**Validation Set**

The noisy and reconstructed images from the validation set are shown in Fig 4a and 4b. The performance evaluation of the PINNED model was done on validation set using the same four evaluation metric defined before.

   i.  **Peak Signal to Noise Ratio (PSNR)**

Across the validation set, PSNR decreases progressively with increasing noise levels, reflecting the increasing difficulty in recovering fine spatial details under elevated noise intensities. *BaTiO3* consistently exhibits the highest PSNR values across all noise levels, indicating that its smooth spatial features allow the model to recover the structural content more effectively. In contrast, *Al* and *ZnMgTe* show a more pronounced decline in PSNR, particularly at higher noise levels, indicating the model's limitations in restoring finer details in these cases.

   ii.  **Structural Similarity Index (SSIM)**

SSIM values decrease systematically with increasing noise levels for all samples, demonstrating the model's challenge to maintain perceptual consistency. Notably, *BaTiO3* retains higher SSIM values, even at severe noise levels, showcasing its relatively consistent structural recovery. Conversely, *ZnMgTe* demonstrates a steep SSIM decline, indicating significant loss of structural information under noisy conditions. *Al* occupies an intermediate position, balancing structural preservation across noise levels.

   iii.  **Fourier Spectrum Consistency (FSC)**

For the validation set, *BaTiO3* maintains relatively stable FSC values across lower noise levels, highlighting the model's ability to preserve frequency details in smooth-structured images. However, at higher noise intensities (Levels 4 and 5), *ZnMgTe* exhibits a marked reduction in FSC, indicating substantial perturbations in frequency-domain information. *Al* remains moderately consistent, showing gradual decline across increasing noise levels.

   iv.  **Edge Preservation Index (EPI)**

EPI values show a gradual upward trend with increasing noise levels, suggesting the model's increased difficulty in accurately preserving edges as the noise becomes more severe. Among the validation samples, *Al* maintains consistently higher EPI values, reflecting its sharper edge content and the model's success in edge recovery. *ZnMgTe*, on the other hand, shows the lowest EPI values, likely due to its complex structural details and texture, which challenge the model's edge preservation capabilities.

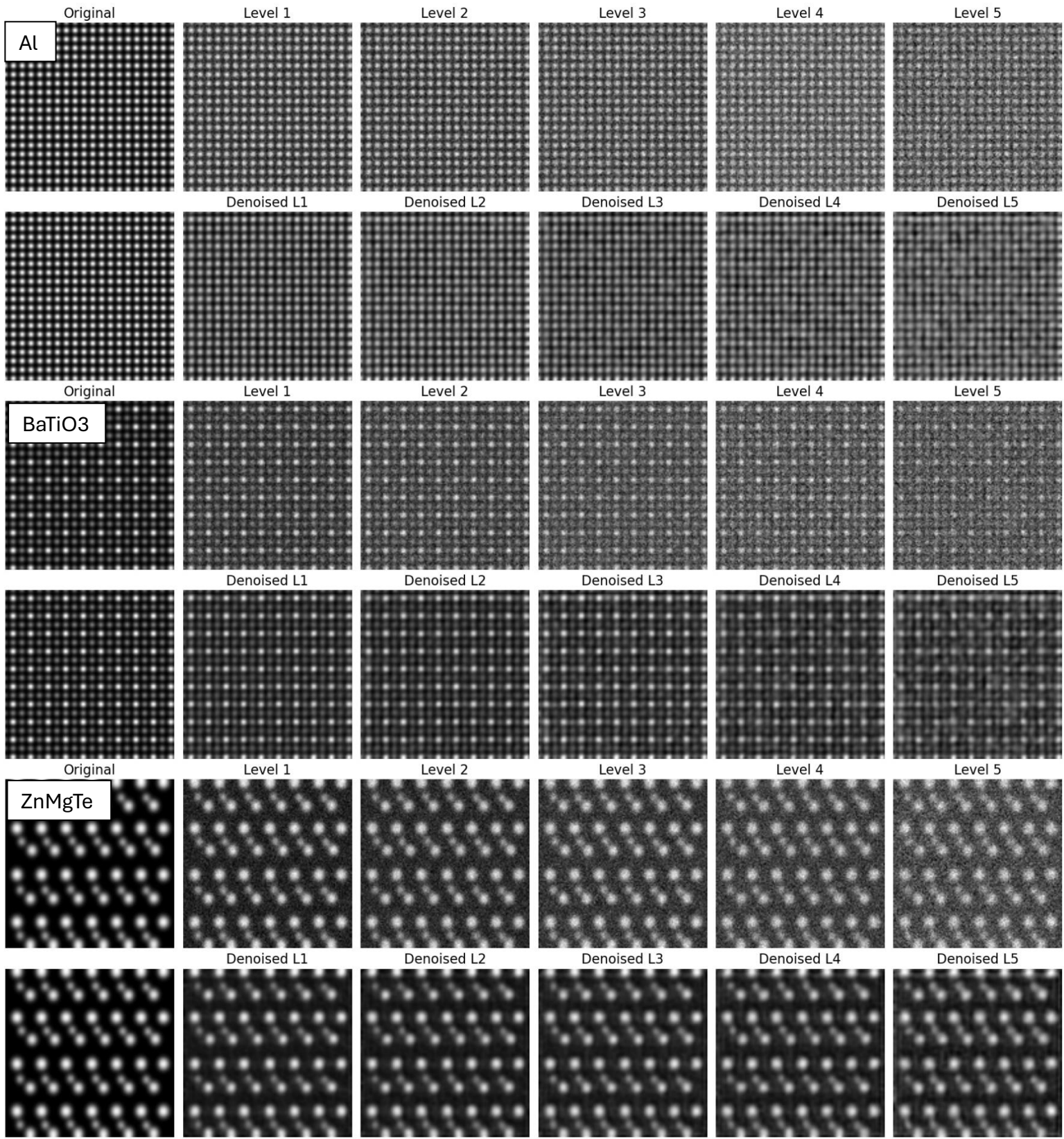

**Figure 4:** Noisy and reconstructed STEM images from validation dataset

*BaTiO3* demonstrates moderate EPI performance, balancing edge recovery and noise robustness.

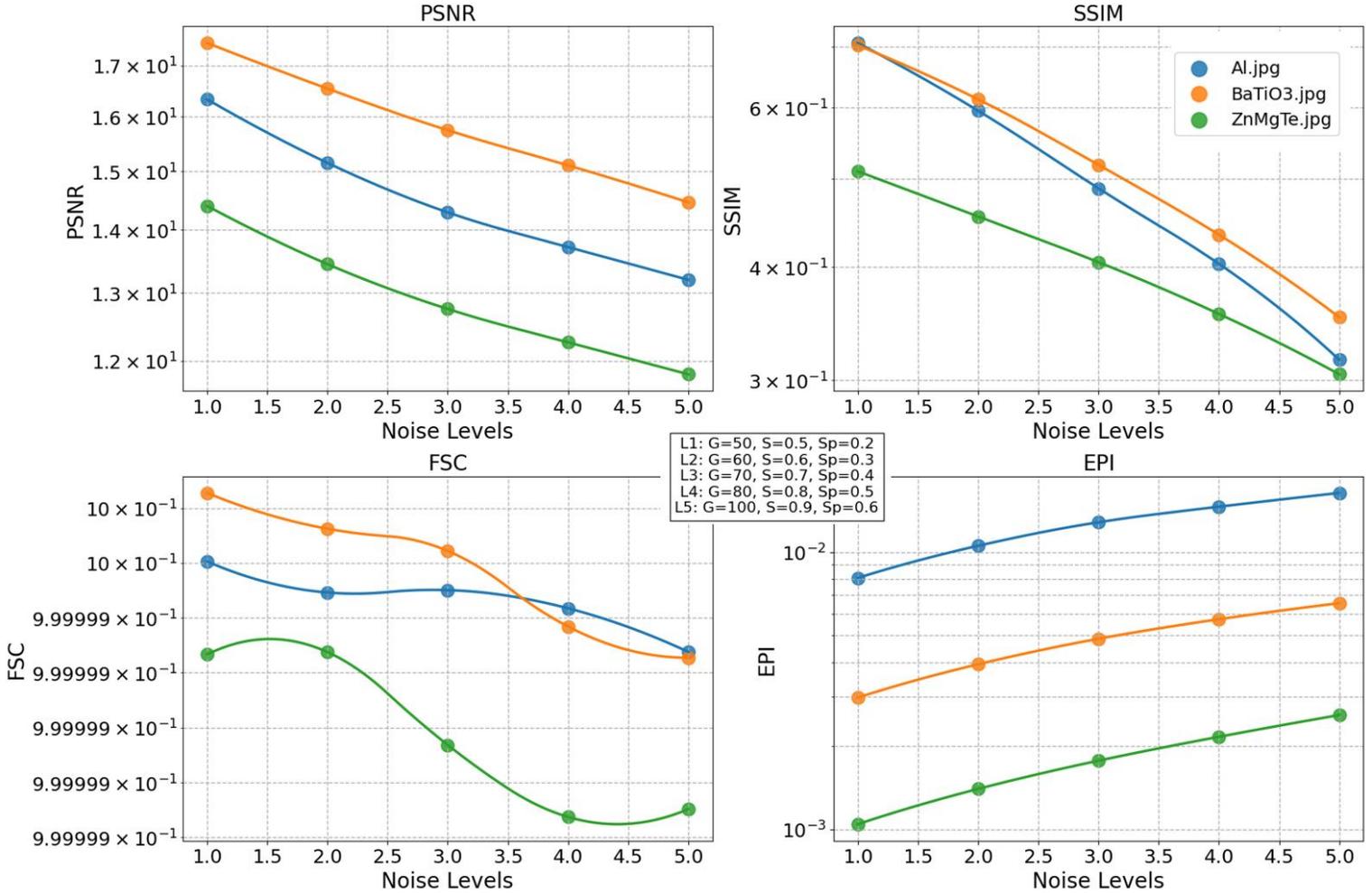

**Figure 5:** Performance evaluation metrics for all images in validation set

**Conclusion**

In this work, we proposed and evaluated PINNED, a physics inspired deep learning framework for denoising STEM images degraded by multi-source noise. The model architecture combines a deeper convolutional encoder decoder structure with targeted loss functions mean squared error (MSE), total variation, spectral fidelity, brightness, and contrast consistency losses to optimize for both pixel level accuracy and structural fidelity.

The evaluation was conducted across synthetic datasets with progressively increasing noise levels, incorporating Gaussian, static, and speckle noise components. Results demonstrated that PINNED achieves robust denoising performance, as evidenced by consistent trends in four quantitative metrics namely peak signal to noise ratio (PSNR), structural similarity index (SSIM), Fourier spectrum consistency (FSC) and Edge perseveration index (EPI).

Overall, the proposed PINNED framework demonstrates strong denoising performance, balancing detail preservation, structural recovery, and spectral consistency across diverse noise conditions. These results underscore the model's potential for improving the quality of STEM images, enabling enhanced analysis in scientific and industrial applications. Future work will explore further optimization of the loss functions and the integration of domain-specific physics constraints to enhance the robustness of PINNED for real world imaging challenges.